\documentclass{pasj00}

\begin{document}
\SetRunningHead{Nomura et al.}{Line-driven disk wind model for BAL QSOs} 
\Received{2012/10/11}
\Accepted{2012/11/14}

\title{Modeling line-driven disk wind for
broad absorption lines of quasars}

\author
{Mariko \textsc{Nomura}\altaffilmark{1},
Ken \textsc{Ohsuga}\altaffilmark{2, 3},
Keiichi \textsc{Wada}\altaffilmark{4},
Hajime \textsc{Susa}\altaffilmark{5}, 
and 
Toru \textsc{Misawa}\altaffilmark{2,6}
}

\altaffiltext{1}{Department of Physics, Ochanomizu University, 2-1-1 Otsuka, Bunkyo, Tokyo 112-8610}
\altaffiltext{2}{National Astronomical Observatory of Japan, Osawa, Mitaka, Tokyo 181-8588}
\altaffiltext{3}{School of Physical Sciences, Graduate University of
Advanced Study (SOKENDAI), Shonan Village, Hayama, Kanagawa 240-0193}
\altaffiltext{4}{Graduate School of Science and Engineering, Kagoshima University, Kagoshima 890-0065}
\altaffiltext{5}{Department of Physics, Konan University, 8-9-1 Okamoto, Higashinadaku, Kobe}
\altaffiltext{6}{School of General Education, Shinshu University, 3-1-1 Asahi, Matsumoto, Nagano 390-8621}
\email{mariko@cosmos.phys.ocha.ac.jp, ken.ohsuga@nao.ac.jp, wada@astrophysics.jp, susa@konan-u.ac.jp, misawatr@shinshu-u.ac.jp}


%


\maketitle

\begin{abstract}
The disk wind, which is powered by the radiation force due to spectral lines (line force), 
is studied for broad absorption line (BAL) quasars. 
We investigate the structure of the disk wind based on the non-hydrodynamic method and compare with
wind properties inferred from X-ray observations of BAL quasars.
In this paper, we apply the stellar wind theory to the initial condition 
(the mass outflow rate at the base of the wind).
We found the funnel-shaped winds 
with a half opening angle of $\sim 50^{\circ}$ 
for the case of $\epsilon=0.3-0.9$ and $M_{\rm BH}=10^{7-8.5}M_\odot$,
where $\epsilon$ is the Eddington ratio 
and $M_{\rm BH}$ is the black hole mass.
Thus, the absorption features are observed 
for an observer of which a viewing angle is around $50^{\circ}$.
A probability of BAL quasars
is $\sim 7-11\%$, which is roughly consistent 
the abundance ratio of BAL quasars, $\sim 10-15\%$.
Here, the probability is estimated by
the solid angle, that the absorbing features would be detected,
divided by $4\pi$.
In contrast, if the Eddington ratio is smaller than $0.01$
or if the black hole is very massive, $M_{\rm BH}\gtrsim 10^9M_{\odot}$,
the disk wind is not launched due to the less effective line force.
Then, the quasars are identified as non-BAL quasars
independently of the observer's viewing angle.
\end{abstract}

\section{Introduction}
Accretion disks surrounding massive black holes 
are origins of activity of active galactic nuclei (AGNs).
Their continuum emission comes from the disks
and absorption as well as emission lines are thought to be 
produced by the matter above and/or around the disks.
%
Broad absorption lines (BALs),
of which line widths exceed $2,000\,\rm{km\, s}^{-1}$,
are observed in $\sim 10-20\%$
of quasars \citep{We91, Homann93, Allen11}.
The BALs are caused by metals 
moderately ionized
and 
blueshifted with typical speeds of $\sim 10,000\, \rm{km\,s}^{-1}$, up to 
$\sim 0.2c$ or more (e.g., \cite{J96}).
\citet{We91} suggested that
the difference between the BAL and non-BAL quasars 
is caused by observers' viewing angles,
since the properties of emission lines and continua 
of the BAL and non-BAL quasars are remarkably similar.
\citet{Elvis00} proposed a phenomenological 
model of funnel-shaped disk wind,
in which the matter is blown away toward the 
direction of the polar angle of $\sim 60-66^{\circ}$,
to explain the abundance ratio of BAL quasars ($\sim 10\%$)
via the viewing angle.
If outflows launched from the surface of the accretion disks
are accelerated toward the observer, 
the blueshift of the BALs could be explained.
On the other hand, 
if the wind does not interrupt the line of sight,
then BAL features would not emerge in the spectra.

Many theoretical models have been proposed to account for the 
origin of the outflows so far.
One plausible scenario is that of `magnetically driven winds' 
(\authorcite{B82} \yearcite{B82};
\authorcite{K94} \yearcite{K94};
\authorcite{Ever07} \yearcite{Ever07}).
In this model, the matter is magnetically accelerated.
However, this model needs an extra mechanism 
to explain that the metals are moderately ionized,
since the gas irradiated by the strong X-ray around the nucleus
is fully photoionized
(the so-called overionization problem).

Another plausible force that accelerates the disk wind 
is the radiation force due to spectral lines (line force).
This model can explain both acceleration and ionization states.
In this model, the gas on the surface of the accretion disk is accelerated 
by absorbing ultraviolet (UV) 
radiation from the disk through the bound-bound transition. 
Since the bound-bound absorption does not 
effectively occur for the overionized metals,
the line force can accelerate the gas only in the lower-ionization state.
Indeed, \citet{SK90} showed that the line force 
works efficiently for the gas in the lower-ionization state
and could be much more powerful than the radiation force 
due to electron scattering.
Thus, though the luminosity of the disks in most quasars
does not exceed the Eddington luminosity,
the line-driven disk wind can be launched.
The discovery of the line locking in the quasars
\citep{F87} strongly supports the idea that 
the line force plays an important role in driving the outflows.
\authorcite{Proga98} (\yearcite{Proga98}, \yearcite{Proga99}) performed 
two-dimensional radiation-hydrodynamic (RHD) simulations 
of winds from disks around white dwarfs.
This RHD method has been improved 
and applied to the disk winds in AGNs
by \citet{Proga00} and \citet{Proga04} (hereafter PK04),
in which the radiation transport of 
X-ray and UV is taken into consideration.
They showed 
the funnel-shaped disk winds
of which the opening angle is $\theta \sim 70^\circ$
where $\theta$ is the polar angle measured from the rotation axis.
The line-driven wind is launched 
at the distance from the black hole of $\sim$ several $100 R_{\rm S}$,
where $R_{\rm S}$ is the Schwarzschild radius,
and the ejected matter goes away mainly in the 
direction of $\theta \sim 70^{\circ}$.
Their simulations revealed that the line-driven wind is in fact formed 
for a set of the black hole mass ($M_{\rm{BH}}=10^8M_{\odot}$) 
and the Eddington ratio ($\epsilon=0.5$).
PK04 suggested that winds can be produced for $M_{\rm BH}>10^7 M_{\odot}$, 
but no winds appear for $\epsilon=0.1$. 
This parameter dependence should be explored for a wider range of 
the parameters. 
\citet{Schurch09} and \citet{Sim10} calculated the spectra and compare them with those of the X-ray observations. 
However, they also only investigated the case of $\epsilon=0.5$ and $M_{\rm BH}=10^8M_{\odot}$.


\citet{Risaliti10} studied the disk winds in AGNs
for the wide parameter space 
of the black hole mass and the Eddington ratio.
They investigated steady structures 
by non-hydrodynamic calculations,
in which they solved the trajectories of the matter ejected 
from the disk surface without 
calculating the RHD equations.
Their results seem to be consistent with 
the results of the RHD simulations.
This non-hydrodynamic method is a powerful tool 
to investigate the dependence on several unknown parameters.
Here, we use a similar method to investigate BAL quasars.
However, \citet{Risaliti10} did not research the wind properties, 
the ionization parameter, the velocity and the column density along the line of sight. 
Thus, within the framework of the non-hydrodynamic model,
it is worth studying whether the disk wind
explains the BAL features.

In \citet{Risaliti10} and PK04,
the density and the velocity 
at the base of the wind (at the disk surface)
are treated as free parameters
and simply assumed to be independent of the 
distance from the black hole.
However, the stellar wind driven by the line force
was investigated in detail.
CAK75 derived the mass outflow rate (the density and the velocity)
as functions of the gravity and the radiative flux.
Thus, we should employ the mass outflow rate of the wind base 
consistent with the prediction of CAK75.
Based on the CAK theory, 
the density at the wind base
is not constant but decrease with distance 
from the black hole,
if the velocity is assumed to be the sound velocity
(see section \ref{sec:basic-eq}).

In the present paper,
employing the wide range of the parameters
of the black hole mass, 
the Eddington ratio,
we investigate the conditions 
under which the disk wind model can reproduce the 
X-ray absorption features of the BAL quasars.
Although we do not study the spectra, we calculate the ionization parameter, 
the velocity and the column density for wide $\epsilon$-$M_{\rm{BH}}$ range 
and compare them with those inferred by the X-ray observation. 
We describe our calculation method in \S\,2.
We present results in \S\,3.
Finally, \S\,4 and \S\,5 are devoted to discussion and conclusions.

\section{Numerical Methods and Models}
\subsection{Outline}
We study a steady structure of line-driven disk winds
and apply them to the BAL quasars.
We investigate the conditions 
under which the disk wind model can explain the 
X-ray absorption features seen in the BAL quasars.

The method for investigating the structure of the wind
is basically the same used by \citet{Risaliti10},
but, based on the stellar wind theory, 
we employ a more realistic mass outflow rate
proposed by \citet{CAK75} (hereafter CAK75).
Furthermore,
we perform higher resolution calculations
and carefully treat the gas temperature and the ionization state
to compare the model with wind properties inferred from the observations.

\begin{figure}
  \begin{center}
   \FigureFile(80mm,50mm){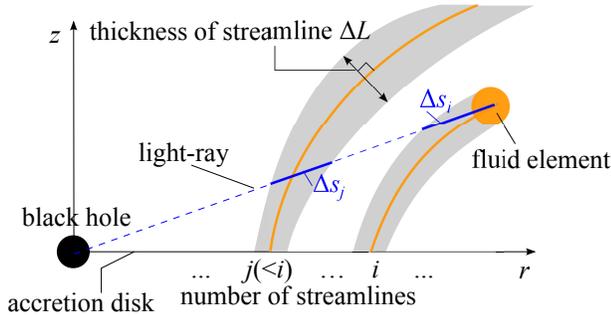}
  \end{center}
 \caption{
Schematic picture for our calculation method.
The black hole and the accretion disk are 
located at the origin and the equatorial plane ($z=0$).
Solid orange lines indicate the streamlines,
and $\Delta L$ is the geometrical thickness of the each flow.
The length of the light ray 
passing through the $j$-th flow
is $\Delta s_j$.
When we calculate the $i$-th streamline,
the extinction of the radiation by 
inner streamlines (1st,2nd,...$i-1$th)
and by the self-shielding of the $i$-th flow
is taken inoto consideration.
 \label{image}
}
\end{figure}

The steady structure of the disk winds is given 
by the calculation of the trajectories of fluid elements 
(streamlines),
which are ejected from the surface
of the geometrically thin and optically thick disks
\citep{SS73}.
We solve the equation of motion considering the line force,
coupling with the mass conservation along the streamline.
We also calculate the temperature 
and the ionization parameter 
by taking account of the radiation transport of 
the X-ray from the vicinity of the black hole 
and UV emitted at the inner part of the disk.
We show the schematic picture for our method in Figure \ref{image}.
We calculate the streamlines 
in order of the distance from the black hole.
When we solve $i$-th streamline, 
we take into consideration 
the extinction 
of the radiation via every inner streamline of $j$-th
($j<i$) as well as
the self-shielding by $i$-th flow.
We can obtain the global structure of the disk winds
after calculating the final (outermost) streamline.

Next, we calculate the column density, the outward velocity,
and the ionization parameter 
for a wide range of viewing angles
and compare them with those inferred from X-ray observation of the BAL quasars.
We evaluate the range of the viewing angle
(i.e., the solid angle),
at which our wind model is consistent with the observations.
The solid angle divided by $4\pi$ means the probability 
that the system is identified as a BAL quasar
(hereafter, BAL probability).
We investigate the BAL probability 
for a variety of the black-hole mass
and the Eddington ratio.

\subsection{Initial conditions and basic equations}
\label{sec:basic-eq}
By solving the equation of motion including the radiation force,
we investigate the trajectories (streamlines) of the fluid elements
ejected from the disk surface
in the cylindrical coordinates, $(r, \varphi, z)$.
Here, 
the black holes and the accretion disk 
are located at the origin and the equatorial plane ($z=0$).
We assume the system to be axisymmetric 
with respect to the rotation axis.

\begin{figure}
  \begin{center}
   \FigureFile(80mm,50mm){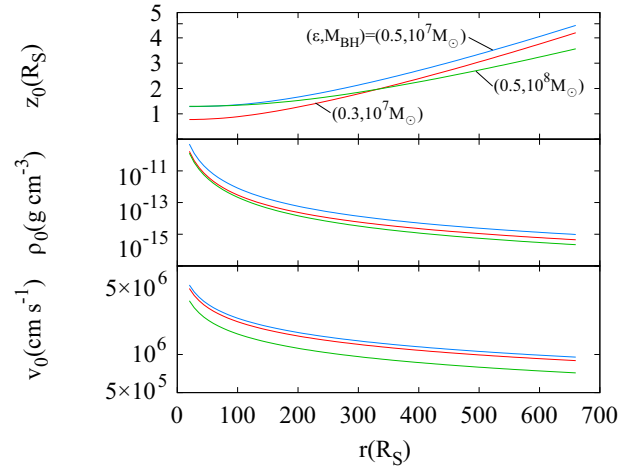}
  \end{center}
 \caption{
 The radial profile of the initial altitude (top panel),
 the initial density (middle panel),
 and the initial upward velocity (bottom panel)
 are ploted 
 for $\epsilon=0.3$ and $M_{\rm BH}=10^7 M_{\odot}$ (red line), 
 for $\epsilon=0.5$ and $M_{\rm BH}=10^7 M_{\odot}$ (blue line),
 and for $\epsilon=0.3$ and $M_{\rm BH}=10^8 M_{\odot}$ (green line).
 \label{init_cond}
}
\end{figure}

The initial position (wind-base) of a fluid element is $(r_0, z_0)$.
The first streamline starts at the innermost radius of 
$r_0=20R_{\rm S}$
(Here, we note that our results do not change so much 
if we employ $r_0=10R_{\rm S}$).
We calculate 81 streamlines
by setting the interval of the initial points in $r$-direction
to be $\Delta r_0=8R_{\rm S}$
(We confirmed
that the wind structure does not change
significantly if we set the interval to be $\Delta r_0=2R_{\rm S}$).
That is, the initial radius of the $i$-th streamline
is $20R_{\rm S}+8(i-1)R_{\rm S}$,
and the final fluid element is launched at the 
radius of $r_0=660R_{\rm S}$.
The initial altitude, $z_0$, is set based on the hydrostatic balance 
in the vertical direction between
the gravity and the pressure forces of gas and radiation,
$GM_{\rm BH}z_0/r_0^3=c_{\rm s}/z_0+(\sigma_{\rm e}/c)\sigma T_{\rm eff}^4$,
where $G$ is the gravitational constant, 
$c_{\rm s}$ is the sound speed of the gas at the mid-plane ($z=0$),
$\sigma_{\rm e}$ is the mass-scattering coefficient for free electrons,  
$c$ is the speed of light,
$\sigma$ is the Stefan-Boltzmann coefficient, 
and $T_{\rm eff}$ is the effective temperature.
In the top panel of Figure \ref{init_cond},
we show $z_0$ as a function of the distance from the black hole, $r$,
normalized by $R_{\rm S}$ being the Schwarzschild radius,
where we employ the alpha viscosity parameter of $0.1$.
We find $z_0$ depends on the black hole mass
and the Eddington ratio (mass accretion rate).

At the wind base ($r=r_0$, $z=z_0$),
we set the horizontal and azimuthal components of the velocity to be
null and the Keplerian velocity.
The initial density, $\rho_0$,
and the vertical component of the initial velocity, $v_0$,
are set so as to meet the prescription by CAK75, in which
they investigated the steady and spherical wind driven by line force.
The resulting mass outflow rate per unit surface at the distance of $R$ is
\begin{eqnarray}
\nonumber  \dot{m}_{\rm CAK}=\frac{1}{\sigma_{\rm e}v_{\rm th}}\frac{GM_*}{R^2}
  \alpha(1-\alpha)^{(1-\alpha)/\alpha} \\
\times (k\Gamma)^{1/\alpha}(1-\Gamma)^{-(1-\alpha)/\alpha},
\end{eqnarray}
where $v_{\rm th}$ is the thermal speed of the gas,
$M_*$ is the mass of the star, 
$\alpha (\sim 0.6)$ and $k (\sim 0.03)$ are 
the constants related to the line force,
$\Gamma$ is the Eddington ratio of the star
[see equation (21) of CAK75].
By making minor revisions,
we apply above relation to the mass outflow rate from the disk surface.
The gravity $GM_*/R^2$ is replaced by $GM_{\rm BH} z_0/r^3$,
which is the gravity at the wind base.
We use the local Eddignton ratio,
\begin{equation}
  \Gamma'=\frac{\sigma_{\rm e}\sigma T_{\rm eff}^4/c}{GM_{\rm BH}z_0/R^3}=\frac{3\epsilon}{4\eta z_0/R_{\rm S}},
\label{eq:gamma}
\end{equation}
as substitute for $\Gamma$.
As a result, we have
\begin{eqnarray}
\nonumber  \rho_0 v_0=\frac{1}{\sigma_{\rm e}v_{\rm th}}\frac{GM_{\rm BH}z_0}{r^3}
  \alpha(1-\alpha)^{(1-\alpha)/\alpha} \\
\times (k\Gamma')^{1/\alpha}(1-\Gamma')^{-(1-\alpha)/\alpha}.
\label{init_rho}
\end{eqnarray}
Throurghout the present study, 
we assume the vertical component of the initial velocity
to be the sound velocity at the disk surface, 
$v_0=(k_{\rm B} T_{\rm eff}/\mu m_{\rm p})^{1/2}$,
where $k_{\rm B}$ is the Boltzmann constant,
$\mu(=0.5)$ is the mean molecular weight,
and $m_{\rm p}$ is the proton mass.
Thus, the initial density, $\rho_0$, is given by the equation
(\ref{init_rho}). 
In Figure \ref{init_cond}, we show $\rho_0$ (middle panel)
and $v_0$ (bottom panel) as a function of $r/R_{\rm S}$.
Since $v_0$ is much smaller than the escape velocity,
the disk wind appears if the fluid elements are
accelerated by the line force.

Here we note that 
the density and velocity of the wind base are still unknown,
although we apply the CAK75 relation.
For instance, 
the mass outflow rate might be enhanced if 
the magnetic pressure force cooperates for
launching the disk wind.
High resolution radiation (magneto) hydrodynamic simulations
would make clear the point.




The equation of motion including the radiation force is
\begin{equation}
  \frac{d \boldsymbol v}{dt}=-\frac{GM_{\rm{BH}}\boldsymbol R}{R^3}+
  \frac{l^2\boldsymbol r}{r^4}
  +\frac{\sigma_{\rm e}(\boldsymbol F_{\rm{UV}}+\boldsymbol F_{\rm{X}})}{c}+M\frac{\sigma_{\rm e}\boldsymbol F_{\rm{UV}}}{c},
  \label{EOM}
\end{equation}
where $\boldsymbol v=(v_{\rm r}, v_{\rm{\varphi}}, v_{\rm z})$ is the velocity, 
$R\equiv[(r^2+z^2)^{1/2}]$ is the distance from the origin,
$\boldsymbol R$ is the point vector of the fluid element,
$l$ is the specific angular momentum, 
$\boldsymbol  F_{\rm X}$ and $\boldsymbol  F_{\rm{UV}}$ are 
the X-ray flux and the UV flux, 
and $M$ is the force multiplier (see below).
The radiation force due to the free electrons
is the third term of the right-hand side.
The final term is
line-driven radiation force,
which may cause the disk wind in our model.
The gas pressure is neglected.

Here, we speculate that the wind flow is launched from the disk surface perpendicularly, and is then directed away from the
central source. 
When the flow is perpendicular to the disk plane, the geometrical thickness of each flow ($\Delta L$, see Figure \ref{image}), 
which is measured at perpendicular to the streamline on $r$-$z$ plane, 
is kept constant, $\Delta L = \Delta r_0$. 
If we simply assume $\Delta L\propto r$ when the wind is blown outward, 
the relation of $\Delta L=(r/r_0)\Delta r_0$ can describe the geometrical thickness of the flow both 
in the vertical and outward phases. 
Then, the density is given from the mass conservation equation ($2\pi r\Delta L\rho |v|=\rm{const}$) as 
\begin{equation}
 \rho (v_r^2+v_z^2)^{1/2} r^2=\rm const.
 \label{dens_1}
\end{equation}

For comparison, we employ the other model, in which we assume the width of the flow does not broaden (the width is constant) 
not only when the flow is vertical to the disk, but also 
when the flow is blown outward; $\Delta L = \Delta r_0$. 
In this case, the density is estimated by $\rho (v_r^2+v_z^2)^{1/2} r=\rm const$. 
%
In the latter model, the density of the wind matter is relatively large 
at the distant region ($r>>r_0$). 
Thus, the maximum of the ionization parameter becomes smaller. 
However, since the large fraction of the moderately ionized region appears 
where the flow is vertical to the disk ($r \sim r_0$), 
the BAL probability (see the \S\,\ref{BALC}) is nearly independent of the model of the wind geometry. 

\subsection{Radiative fluxes of X-ray and Ultraviolet}

%
%
%

We consider two spectral components of the radiation, X-ray and UV. 
The luminosities of X-ray ($L_{\rm X}$) and UV ($L_{\rm UV}$) 
are given by $L_{\rm X}=f_{\rm X} \epsilon L_{\rm edd}$
and $L_{\rm UV}=(1-f_{\rm X}) \epsilon L_{\rm edd}$.
Here, $f_{\rm X}$ is the X-ratio, $L_{\rm X}/(L_{\rm X}+L_{\rm UV})$,
$\epsilon$ is defined as
$\epsilon\equiv (L_{X}+L_{\rm UV})/L_{\rm edd}$,
and $L_{\rm edd}$ is the Eddington luminosity.
In the present study, 
$\epsilon$ is a parameter and we employ $f_{\rm X}=0.15$.

X-ray is assumed to be emitted near the central black hole 
and treated as a central point source.
The radial component ($R$-component) of the X-ray flux is written as
\begin{equation}
 F_{\rm X}^R=\frac{L_{\rm X}}{4\pi R^2} e^{-\tau_{\rm X}},
\end{equation}
where 
$\tau_{\rm X}$ is the optical depth for X-ray measured 
from the origin.
%
The other components of the X-ray flux are zero.
We consider that the radiation is attenuated by the wind flow
in the present study. 
When we calculate the $i$-th streamline,
we take into consideration the obscuration 
by every inner streamline of $j$-th ($j<i$).
That is, if a ray from the center to
the fluid element gets across the 
$j$-th streamline ($j<i$),
the radiation suffers from the dilution.
Even if the ray does not get across, 
the $j$-th flow sometimes contributes to the obscuration
because the flow has geometrical thickness (see \S\,\ref{sec:basic-eq}).
We also consider the self-shielding effect,
which is extinction caused by the upper stream of the $i$-th flow.
Therefore, the optical depth is calculated by
\begin{equation}
 \tau_{\rm X}=\sum_{j\leq i}\Delta \tau_{{\rm X},j},
  \label{tau_N}
\end{equation}
where $\Delta \tau_{{\rm X}, j}=\rho\sigma_{\rm X}\Delta s_j$,
with $\sigma_{\rm X}$ being the mass extinction coefficient for X-ray
and $\Delta s_j$ being the length that the ray 
passing through the $j$-th flow (see Figure \ref{image}).
The mass extinction coefficient
is defined by $\sigma_{\rm X}=\sigma_{\rm e}$ for $\xi\ge 10^5$
and $\sigma_{\rm X}=100 \sigma_{\rm e}$ for $\xi< 10^5$,
where $\xi$ is the ionization parameter defined as,
\begin{equation}
 \xi{\rm\,[erg\, cm\,s^{-1}]}
  =\frac{m_{\rm p} L_{\rm X}}{\rho R^2} e^{-\tau_{\rm X}}.
  \label{xi}
\end{equation}
This simple treatment of $\sigma_{\rm X}$ reflects
that the photoelectronic absorption is not effective
when the ionization parameter is very high,
since the number of bound electrons is too small 
(see also \cite{Risaliti10}).
We have the $r$- and $z$-components of the X-ray flux is
$F_{\rm X}^r=F_{\rm X}^R (r/R)$ and
$F_{\rm X}^z=F_{\rm X}^R (z/R)$.
The azimuthal component is zero.

We suppose that 
an optically thick and geometrically thin accretion disk
is the only UV source
where the effective temperature follows
the radial profile of 
\begin{equation}
 T_{\rm eff}=T_{\rm in} \left( \frac{r}{r_{\rm in}} \right)^{-3/4},
\end{equation}
where $r_{\rm in}(=3R_{\rm S})$ is the disk inner radius
and $T_{\rm in}$ is the effective temperature at $r=r_{\rm in}$.
We set $T_{\rm in}$ so as to meet the condition of
\begin{equation}
 (1-f_{\rm X})\epsilon L_{\rm edd}=\int_{r_{\rm in}}^{r_{\rm out}}
 2\pi r \sigma T_{\rm eff}^4 dr,
\end{equation}
and the UV flux from the disk 
in the optically thin media is calculated by 
\begin{equation}
  \boldsymbol F_{\rm thin}=\int \frac{\sigma T_{\rm eff}^4}{\pi}
  \boldsymbol n d\Omega,
  \label{UVflux}
\end{equation}
where $\sigma$ is the Stefan-Boltzmann coefficient,
$\boldsymbol n$ is the unit vector,
and $\Omega$ is the solid angle.
Here, note that we integrate the radiation of the disk 
within $r=r_{\rm out}$, at which 
the effective temperature of the disk is $10^4 \,\rm K$,
since the outer part is too cold to emit UV photons effectively,
and since the disk luminosity of the outer part is 
much smaller than that of the inner part.
For simplicity, 
we attenuate the radial component ($R$-component) of the UV flux 
using the optical depth measured form the origin, 
$\tau_{\rm UV}$, as 
\begin{equation}
 F_{\rm UV}^R=F_{\rm thin}^R e^{-\tau_{\rm UV}}.
\label{UVR}
\end{equation}
The numerical method for calculating $\tau_{\rm UV}$ is 
the same as that for X-ray optical depth,
but we use $\sigma_{\rm e}$ instead of $\sigma_{\rm X}$.
We assume that 
the dilution of the polar component of the UV flux 
is negligible,
\begin{equation}
 F_{\rm UV}^{\theta}=F_{\rm thin}^{\theta},
\label{UVtheta}
\end{equation}
since the wind is expected to flow near the disk surface.
The azimuthal component of the flux is zero by definition.
In numerically, 
we divide the surface of the accretion disk into $256 \times 256$ small surface elements
and calculate the UV flux.

\subsection{Force multiplier}
In order to estimate the line force,
we employ the force multiplier proposed by 
\citet{SK90}, which is modified version of CAK75.
The force multiplier,
which is ratio of line force to electron-scattering force,
is written as 
\begin{equation}
  M(t,\xi)=kt^{-0.6}\Biggl[ \frac{(1+t\eta_{\rm{max}})^{0.4}-1}{(t\eta_{\rm{max}})^{0.4}}\Biggr],
  \label{forceM}
\end{equation}
Here, $t$ is the local optical depth parameter
\begin{equation}
  t=\sigma_{\rm e} \rho v_{\rm{th}}\Bigl| \frac{dv}{ds}\Bigr|^{-1},
   \label{t-xi}
\end{equation}
where $v_{\rm{th}}=(k_{\rm B} T/\mu m_{\rm p})^{1/2}$ 
is thermal velocity with $T$ being the gas temperature,
and $dv/ds$ is the velocity gradient along the light-ray.
Since $k$ and $\eta_{\rm{max}}$ depend on $\xi$ as,
\begin{equation}
  k=0.03+0.385\exp(-1.4\xi^{0.6}),
\end{equation}
and 
\begin{equation}
 \log_{10}\eta_{\rm{max}}=\left\{ 
    \begin{array}{ll}
      6.9\exp(0.16\xi ^{0.4}) & \log\xi \le 0.5 \\
      9.1\exp(-7.96\times 10^{-3}\xi) & \log\xi >0.5 \\
    \end{array} \right.
 ,
  \label{k_eta}
\end{equation}
the force multiplier is the function of 
the local optical depth parameter, $t$,
and the ionization parameter, $\xi$. 

In our calculation, 
we approximate the velocity gradient 
\begin{equation}
 \frac{dv}{ds}=\frac{dv_{\rm l}}{dl},
  \label{grad_v}
\end{equation}
where $dv_{\rm l}/dl$ is the 
velocity gradient along the streamlines.
As we will show in Figure 3, 
the fluid element is lifted up in nearly vertical direction, since
the vertical radiation flux from around the wind base 
mainly accelerate the matter upwards.
Thus, the velocity gradient along the right-ray would be 
approximated by that along the streamlines.
Subsequently, 
the matter is blown away in the radial direction.
Then, the inner part of the accretion disk 
mainly illuminate the outflowing matter. 
Thus, above approximation would be valid.
We calculate the gas temperature
assuming that the gas is in the radiative equilibrium, 
since the radiative cooling/heating timescale is much smaller 
than the cooling timescale via the adiabatic expansion \citep{Proga00},
\begin{eqnarray}
n^2(G_{\rm{Compton}}+G_{\rm X}-L_{\rm{b,l}})=0,
\label{heat_cool}
\end{eqnarray}
where, $G_{\rm{Compton}}$ is Compton heating/cooling rate,
\begin{eqnarray}
G_{\rm{Compton}}=8.9\times10^{-36}\xi(T_{\rm X}-4T),
\end{eqnarray}
$G_{\rm X}$ is the rate of X-ray photoionization heating and recombination cooling,
\begin{eqnarray}
G_{\rm X}=1.5\times10^{-21}\xi^{1/4}T^{-1/2}(1-T/T_{\rm X}),
\end{eqnarray}
$L_{\rm{b,l}}$ is the bremsstrahlung and line cooling,
\begin{eqnarray}
\nonumber L_{\rm{b,l}}=3.3\times10^{-27}T^{1/2}+1.7\times10^{-18}\xi^{-1}T^{-1/2}\\
\times \exp(-1.3\times10^5/T)+10^{-24}.
\end{eqnarray}
We assume the temperature of the X-ray radiation, 
$T_{\rm X} =10^8 \rm{K}$,
which is the same as \citet{Proga00}.
Since this method for calculating the gas temperature
is almost valid in a
low-density and high-ionization parameter regime,
the gas temperature would be underestimated 
in the region of the high-density and low-ionization parameters.
Thus, we set the lower limit of the gas temperature 
as being equal to the disk effective temperature,
$T(r,z)_{\rm min}=T_{\rm eff}(r)$.

Although, in the present work, the force multiplier depends on the 
gas temperature through the sound speed [see equation (\ref{forceM}) and (\ref{t-xi})],
another model for the line force,
which does not depends on the gas temperature,
has been proposed by \citet{Gayley95}.
Our results does not change so mach 
if we employ another force multiplier,
since 
the gas temperature is kept around $10^4$ K
in the region that the matter is mainly accelerated.

\subsection{Assessment of BAL probability}
\label{BALC}

A probability for observing BAL (BAL probability) is evaluated as follows.
Using the resulting structure of the disk wind,
we investigate the ionization parameter, 
the outward velocity, and the column density
along lines of sight
(from an observer to the central black hole).
The X-ray observations of the BAL quasars reported 
the ionization parameter to be
$\lesssim 500$ 
(\authorcite{Re03} \yearcite{Re03}, \authorcite{Ga04} \yearcite{Ga04})
and ${\rm several}\times 1000$ 
(\authorcite{Br06} \yearcite{Br06}, \authorcite{Wa08} \yearcite{Wa08}).
The X-ray observations also reveal 
that the outward velocity and 
the column density of the gas in such lower-ionization state
are larger than $10^4\,\rm{km\,s}^{-1}$
and larger than $10^{23}\,\rm{cm^{-2}}$
(\authorcite{Ch07} \yearcite{Ch07} and references therein).
Thus, we consider two conditions:
(A)the outward velocity of the matter with $\xi<100$ 
exceeds $10^4\,\rm{km\,s}^{-1}$ and
(B)the column density of the gas with $\xi<100$ 
is larger than $10^{23}\,\rm{cm^{-2}}$.
We also add the condition that 
(C)the column density is smaller than $1.5\times 10^{24}\,\rm{cm^{-2}}$,
since the Compton thick objects would not be identified as BAL quasars.
If three conditions are satisfied,
we suppose that the X-ray absorption features emerge in the spectra.
If this is the case, our wind model could also
explain the BALs in UV band,
since the X-ray absorption closely correlate to the BALs seen
in UV wavelength 
\citep{B00}.
We evaluate the solid angle, $\Omega_{\rm{BAL}}$, 
in which above conditions are satisfied.
The BAL probability is given by $\Omega_{\rm{BAL}}/4\pi$.

\section{Results}
\subsection{Structure of line-driven wind}
\label{results}

\begin{figure}
  \begin{center}
   \FigureFile(65mm,50mm){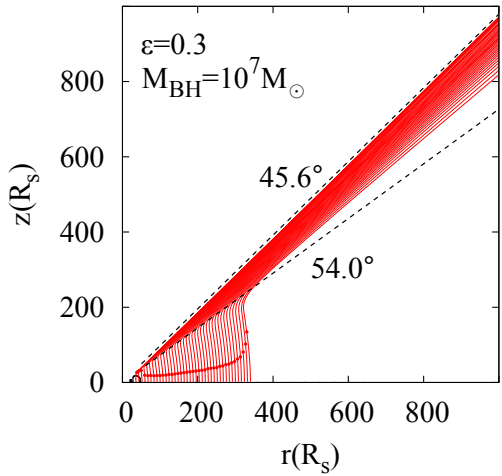}
  \end{center}
 \caption{
 The steady structure of the line-driven disk wind 
 for $\epsilon=0.3$ and $M_{\rm BH}=10^7M_{\odot}$.
 The black hole and the accretion disk are 
 located at the origin and the equatorial plane ($z=0$).
 The red solid lines show trajectories of fluid elements (streamlines) 
 ejected from the surface of the accretion disk.
 The red circles on the red lines indicate the point where
 the wind velocity exceeds the escape velocity.
 At the inner region of $r\lesssim 30R_{\rm S}$,
 the ejected gas immediately goes back to the disk surface
 (black lines).
 It is also found that the wind is not launched at the outer region,
 $r>350R_{\rm S}$, although we can not see the streamlines in this figure.
 The BAL are observed at the viewing angle of
 $\theta=45.6^\circ-54.0^\circ$ (between dotted-lines).
 \label{base}
 }
\end{figure}

In Figure \ref{base}, red and black solid lines 
show trajectories of fluid elements 
(streamlines) ejected from the surface of the accretion disk 
in $r$-$z$ plane. 
The mass of the black hole and the Eddington ratio,
which are parameters in our calculation,
are given as $M_{\rm BH} = 10^7M_{\odot}$ and $\epsilon = 0.3$
(hereafter, we call these 'baseline parameters').
The initial density, velocity, 
and altitude of streamlines are shown 
in Figure \ref{init_cond} (red lines).
On the red lines, 
the matter is accelerated and attain a velocity equal to the escape velocity, 
$v=v_{\rm esc}$ where 
$v=(v_{\rm r}^2+v_{\rm{\varphi}}^2+v_{\rm z}^2)^{1/2}$,
at the points of the red filled circles.
In contrast, the flow velocity does not reach the escape velocity
on the black lines.

In the inner region, $r \lesssim 30R_{\rm S}$, 
it is found that the ejected fluid element immediately returns to
the disk surface (we call this 'failed wind') (see black lines). 
In this region, the initial density is too high 
for the line force to lift up the matter.
High density reduces the ionization parameter
because of $\xi \propto \rho^{-1}$ [see equation (\ref{xi})].
Indeed, we have $\xi$ is much less than $100$ in this region.
In the lower ionization regime ($\xi<100$),
the force multiplier decreases with 
an increase of the density, $M\propto \rho^{-0.6}$,
since $M\propto t^{-0.6}$ where $t \propto \rho$ 
[see the equation (\ref{t-xi})].
Therefore, the line force does not exceed the gravity 
and the gas is not blown away. 
Note that the mechanism of the failed wind 
in our work is different from that in previous works
\citep{Murray95, Risaliti10}. 
We will discuss the point in section \ref{dis-2}.

In the middle region, $30R_{\rm S}\lesssim r \lesssim 350R_{\rm S}$, 
we find that the disk wind is successfully launched.
The matter launched from the disk surface is lifted up
almost perpendicular to the disk plane,
and exceeds the escape velocity (filled circles).
Subsequently, such matter 
is blown away towards the direction with polar angle 
of $\theta \sim 45^{\circ}-55^{\circ}$.
The wind structure is understood as follows.
Since the initial density ($\rho_0 \sim 10^{-14}-10^{-13}$) 
is smaller in this region than in the inner region,
and since the ionization parameter still less than $100$,
the force multiplier is large ($M\propto \rho^{-0.6}$),
typically several $\times 100$.
Thus, the vertical component of the radiation force (line force)
is stronger than the that of the gravity.
In contrast, the radiation from the vicinity of the black hole 
is attenuated by the failed wind, 
thus the radial component of the radiation force
can not play an important role.
As a consequence, the matter is accelerated in the vertical direction.
Such a situation changes at the region of $\theta \gtrsim 55^{\circ}$.
The radiation from the inner accretion disk,
which works to accelerate the matter outward,
increase with a decrease of the polar angle and
is not attenuated by the failed wind in the region.
Therefore, outward radiation force bend the streamlines 
to radial direction and producing the funnel-shaped wind. 

In the outer region beyond $r \sim 350R_{\rm S}$, 
the temperature of the accretion disk is less than $10^4 \rm{K}$ 
and the disk around the wind base does not effectively emit UV photons.
In addition, the UV radiation from the inner region of the accretion disk is 
obscured by the failed wind ($r \lesssim 30R_{\rm S}$) 
and the disk wind ($30R_{\rm S} \lesssim r \lesssim 350R_{\rm S}$).
Thus, although the initial density as well as 
the ionization parameter is small, 
the line force is made powerless and fails to form the wind.


According to the method mentioned in \S\,\ref{BALC}, 
we investigate viewing angles for which the
X-ray absorption features of the BAL quasars are observed. 
In the case of the disk wind with
the baseline parameters, 
the conditions (A), (B), and (C) are satisfied 
at the viewing angle from $\theta=45.6^{\circ}$ to $54.0^{\circ}$
(see \S\,\ref{results}). 
That is, then, the absorption features are observed,
since the lower-ionization matter 
with sufficient radial velocity obscures 
the nucleus (see Figure \ref{base}).
The BAL probability, $\Omega_{\rm BAL}/4\pi$, 
is obtained as $\cos(45.6^{\circ})-\cos(54.0^{\circ}) \sim 11\%$.

For the viewing angle smaller than $45.6^{\circ}$, 
there is no wind between the observer and the nucleus. 
When the viewing angle is between $54.0^{\circ}$ and $63.0^{\circ}$, 
the column density is too large to detect the absorption features 
($N_{\rm H}>1.5\times 10^{24} \rm{cm}^{-2}$),
since the failed wind, of which the density is very high, 
obscures the nucleus.
Thus, although the conditions (A) and (B) are 
satisfied, the object is not identified as the BAL quasars.
For large viewing angle , $\theta>63.0^{\circ}$, 
neither the condition (A) nor (C) is satisfied.
\subsection{BAL Probability}
\label{sec:edd-BH}

\begin{table}
  \begin{center}
    \caption{The BAL probability (\%)}
    \begin{tabular}{cc|cccccc}
      \hline
    & & \multicolumn{6}{c}{Black hole mass ($M_{\rm BH}/M_{\odot}$)} \\
     &   & $10^7$ & $10^{7.5}$ & $10^8$ & $10^{8.5}$ &  $10^9$ & $10^{9.5}$\\
      \hline
              & $ 0.9$  & 7  & 7  & 7  & 5  & 0  & 0 \\
Eddington     & $ 0.7$  & 8  & 8  & 7  & 6  & 0  & 0 \\
ratio ($\epsilon$)   & $ 0.5$  & 9  & 9  & 8  & 8  & 0  & 0 \\
              & $ 0.3$  & 11 & 11 & 11 & 9  & 7  & 0 \\
              & $ 0.01$ & 0  & 0  & 0  & 0  & 0  & 0 \\
      \hline
      \multicolumn{8}{@{}l@{}}{\hbox to 0pt{\parbox{85mm}{\footnotesize
          }\hss}}
    \end{tabular}
\label{table}
  \end{center}

\end{table}

\begin{figure*}
   \begin{center}
 \FigureFile(130mm,50mm){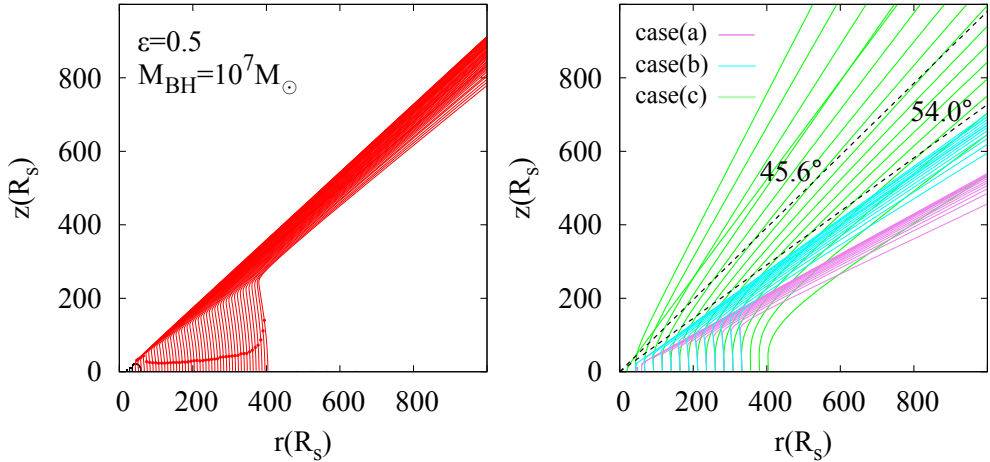}
   \end{center}   
 \caption{
 Same as Figure \ref{base}, but for $\epsilon=0.5$ (left panel).
 The disk structure is almost same with Figure \ref{base},
 although the larger Eddington ratio is employed.
 The right panel is also the same as Figure \ref{base},
 but we employ the initial altitude for $\varepsilon=0.5$ [case (a)],
 the initial density for $\varepsilon=0.5$ [case (b)],
 and the disk luminosity for $\varepsilon=0.5$ [case (c)].
 The larger initial altitude and density work to increase the
 opning angle of the funnel-shaped wind.
 In contrast, the larger luminosity tends to accelerate the matter 
 upward.
}   
\label{edd0503}
\end{figure*}

Table \ref {table} presents the BAL probability. 
For $\varepsilon=0.3-0.9$ and $M_{\rm BH}=10^{7-8.5}M_\odot$,
we find that the BAL probabilities are $5-11\%$
and are not so sensitive to the black hole mass as well as the Eddington ratio.
On the other hand,
the BAL quasars are not observed (BAL probability is null)
in the case of $M_{\rm BH} \gtrsim 10^{9}M_\odot$ or
$\varepsilon\lesssim 0.01$.
%
%

\subsubsection{Eddington ratio dependence}



For $\varepsilon=0.01$, 
the radiation is too weak to accelerate the matter.
Since the wind is not successfully launched,
the absorption features are not observed.
The reason why the BAL probability is roughly kept constant
for $\varepsilon=0.3-0.9$ and $M_{\rm BH}=10^{7-8.5}M_\odot$
is understood by 
the left panel in Figure \ref{edd0503}.
This panel is the same as the Figure \ref{base}, but for $\varepsilon=0.5$.
It is found that the wind launching region broaden as an
increase with $\varepsilon$.
However, we find 
that the opening angle and the thickness of the funnel-shaped wind
does not change so much.
Therefore, the BAL probability is roughly independent of the 
Eddington ratio.

Since the initial values and the disk luminosity
changes with $\varepsilon$,
here, we at first consider an influence of the initial density 
on the wind structure.
For this purpose, 
we employ the initial density profiles for $\varepsilon=0.5$
(blue line of the middle panel in Figure \ref{init_cond}),
but other initial conditions and the disk luminosity
are set to the values for $\varepsilon=0.3$
(red lines of the top and bottom panels in Figure \ref{init_cond}).
The result is plotted by purple lines in the right panel of 
Figure \ref{edd0503} [case (a)].
We find that the resulting disk wind has an wide opening angle
($\theta\sim 65^{\circ}$).
Since the initial density increases as an increase of $\varepsilon$ 
(see Figure \ref{init_cond}),
and since the line force is less effective for the gas with large density
($M\propto \rho^{-0.6}$, see \S \ref{results}),
the height of the failed wind is small
and the stream lines of the disk wind bends 
at the small altitude for the case (a)
in comparison with the case with baseline parameter.
We conclude that the higher density 
tends to make the opening angle of the disk wind larger.

Next, we consider an effect of the initial altitude, $z_0$.
The blue lines of the right panel of Figure \ref{edd0503}
show the result for the case (b) 
in which the initial altitude for $\epsilon=0.5$
(blue line of the top panel in Figure \ref{init_cond})
is supposed, 
although the luminosity, initial density and velocity 
remain the values for $\epsilon=0.3$
(red lines of middle and bottom panels in Figure \ref{init_cond}). 
That is, the larger $z_0$ is employed 
in comparison with the case with the baseline parameters.
Then, since the initial position is distant from the disk surface,
the radiation force becomes less effective relatively and 
the disk wind with the large opening angle is generated.
As we have discussed above, 
the less effective radiation force works to form the lower failed wind,
leading to the disk wind with larger opening angle.


The large disk luminosity tends to narrow the opening angle.
The green lines [case (c)] show the streamlines when the radiation fields
are calculated by assuming $\epsilon=0.5$ 
and other initial conditions remain the values for $\epsilon=0.3$
(red lines in Figure \ref{init_cond}).
Comparing with the streamlines for $\epsilon=0.3$ (Figure \ref{base}), 
the matter, which is launching at $r<150R_{\rm S}$,
is blown away more upward,
inducing the small opening angle of the funnel-shaped wind.
This is because the UV radiation is strong and
the line force effectively accelerates the matter.
We also find that the wind is launching from wide region of the disk,
$r<400R_{\rm S}$,
since the UV emitting region ($T_{\rm eff}\geq 10^4 \rm{K} $) 
expands with an increase of $\varepsilon$.

To sum up, the increase of the luminosity,
which contributes to reduce the opening angle of the wind,
counteracts the increase of the initial density and altitude,
which works to expand the funnel.
As a result,
the BAL probability is insensitive to the Eddington ratio
for $\varepsilon=0.3-0.9$ and $M_{\rm BH}=10^{7-8.5}M_\odot$. 
Here, we note that the initial velocity has little influence 
on the wind structure.
Indeed, 
we verified that the wind structure does not change
if the initial velocity for $\varepsilon=0.5-0.9$ is employed.



\subsubsection{Black hole mass dependence}

\begin{figure}
 \begin{center}
  \FigureFile(65mm,50mm){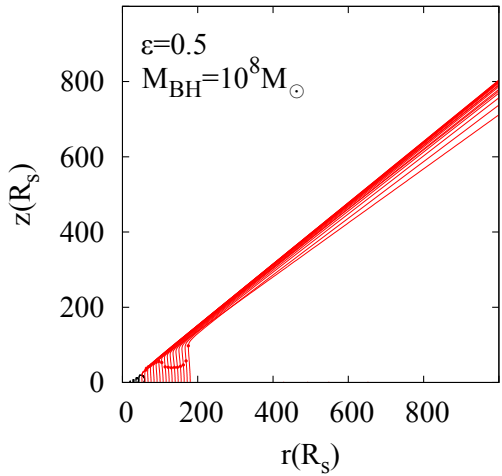}
\end{center}
 \caption{
 Same as the left panel of Figure \ref{edd0503},
 but for $M_{\rm{BH}}=10^8M_{\odot}$.
 The half opening angle of the wind 
 does not depend on the black hole mass.
 However, the wind launching region shrinks 
 with an increase of the black hole mass.
 \label{BH8}
}   
\end{figure}

The BAL probability is insensitive to the 
black hole mass for $M_{\rm BH}=10^{7-8.5}M_\odot$,
but tends to be null for $M_{\rm BH} \gtrsim 10^{9}M_\odot$.
Figure \ref{BH8} shows the streamlines for 
$M_{\rm BH}=10^8M_{\odot}$ and $\epsilon=0.5$.
This figure shows that the opening angle for $M_{\rm BH}=10^8M_{\odot}$ 
is comparable to or slightly smaller than that for 
$M_{\rm BH}=10^7M_{\odot}$.
We also find that 
the launching region,
which is measured in an unit of $R_{\rm S}$,
shrinks with an increase of the black hole mass.
Such behavior is caused by 
the decrease of the force multiplier as well as 
the temperature of the disk.

The local optical depth parameter,
$t=\sigma_{\rm e}v_{\rm th}\rho(dv_l/dl)^{-1}$,
is roughly evaluated as 
$t=\sigma_{\rm e}\rho_0 z_0$ near the wind base,
where we suppose $v_{\rm th}\sim v_0$, $\rho\sim \rho_0$, 
and $dv_l/dl\sim v_0/z_0$.
Since $z_0$ is roughly proportional to $M_{\rm BH}$
($z_0 \propto R_{\rm S}$),
$\Gamma'$ is almost independent form the black hole mass
(see equation \ref{eq:gamma}).
We have 
$\rho_0 \propto (1/v_0^2)(M_{\rm BH}^2/r^3) \propto M_{\rm BH}^{-3/4}$ 
with using the relation of $v_0 \propto M_{\rm BH}^{-1/8}$.
Thus, the optical depth parameter is proportional to
$M_{\rm BH}^{1/4}$, and we have a very weak $M_{\rm BH}$-dependence of 
the force multiplier, $M\propto t^{-0.6} \propto M_{\rm BH}^{-0.15}$. 
As we have discussed above, 
the disk wind with the wide opening angle form 
since the force multiplier is small
(radiation force is less effective).
Thus, the opening angle for $M_{\rm BH}=10^8M_\odot$
is comparable to or 
slightly larger than that for $M_{\rm BH}=10^7M_\odot$.

The size of the UV emitting region 
of the accretion disk ($T_{\rm eff}>10^4 \rm{K}$),
which is normalized by $R_{\rm S}$,
is proportional to $M_{\rm BH}^{-1/3}$,
in the case that the Eddington ratio is kept constant.
This is the main reason that 
the launching region of the wind 
becomes small when the black hole is massive.

Since the UV emitting region becomes narrowed,
and since the failed wind region slightly expands
via the reduction of the line force
($M \propto M_{\rm BH}^{-0.15}$),
the disk wind tends to disappear when the black hole is too massive
($M_{\rm BH} \gtrsim 10^9M_\odot$).
Hence, it is implied that the absorption features are not identified 
in the quasars with very massive black holes.
%
 
\section{Discussion}
\label{dis}
\subsection{Comparison with observations}
Our model indicates that
the BAL probability is several to $10$ percent
for a wide range of the Eddington ratio
and the black hole mass,
$\varepsilon=0.3-0.9$ and $M_{\rm BH}=10^{7-8.5}M_\odot$.
This result is roughly consistent with 
the abundance ratio of BAL quasars, $\sim 10-20\%$
(e.g., \cite{Allen11}).
The resulting BAL probability might increase,
if the nucleus is obscured by the dusty torus, 
which aligns with the accretion disk,
is not identified as quasars.
For instance, 
the BAL probability is doubled, $10-20\%$,
if the observer with $\theta\geq 60^\circ$
cannot detect quasars via the obscuration by the thick torus.
Even if that is the case, 
our result is consistent with the abundance ratio.

In this regime of 
$\varepsilon=0.3-0.9$ and $M_{\rm BH}=10^{7-8.5}M_\odot$,
our model indicates that 
the observer's viewing angle
is responsible for 
the difference between BAL and non-BAL quasars.
If the observer is in the direction of the wind flow
the absorption features are detected.
Otherwise, the nuclei are identified as non-BAL quasars.
That is, our result gives physical bases in phenomenological model
\citep{Elvis00},
in which the dichotomy between the BAL and non-BAL quasars 
is explained by the observer's viewing angle.
Note that, 
in addition to such a viewing-angle effect, 
our results imply that 
the BAL features does not appear 
if the Eddington ratio is too small 
($\varepsilon\lesssim 0.01$)
or if the black holes is too massive 
($M_{\rm BH}\gtrsim 10^{9}M_\odot$).

\begin{figure}
 \begin{center}
  \FigureFile(65mm,50mm){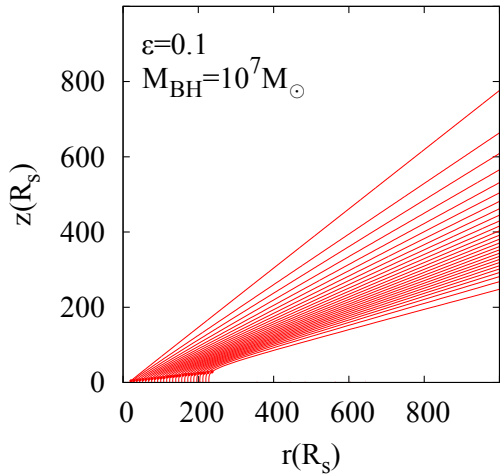}
\end{center}
 \caption{
 Same as Figure \ref{base}, but for $\epsilon=0.1$.
 In this case, the wind with large covering factor forms.
 \label{edd01}
}   
\end{figure}

Here, we note that 
our model overestimate the BAL probability
in the regime of $\varepsilon=0.05-0.1$.
Although our model gives the probability of $20-50$\%,
the accurate value would be much small.
The launching matter is blown away for a wide angle
for $\epsilon=0.05-0.1$, leading to the larger BAL probability.
The streamlines for $\epsilon=0.1$ is plotted in Figure \ref{edd01}.
In this case, the BAL probability is 48 \%.
However, then, the mass outflow rate of the wind 
is a few times larger than the mass accretion rate of the disk.
This implies that our model overproduces the outflows,
and actual BAL probability should be smaller than 48 \%.
In the case that the mass accretion rate is comparable to 
the mass outflow rate, we need to 
solve the outflow and the disk, self-consistently
(we will discuss later).
\subsection{Comparison with previous works}
\label{dis-2}
Throughout the present study, 
we employ the non-hydrodynamic method
that is basically the same with that used in \citet{Risaliti10}.
In \citet{Risaliti10}, 
the initial density and velocity are treated 
as free parameters 
and assumed to be constant in the radial direction,
for simplicity.
However, we employ more realistic profiles of $\rho_0$ and $v_0$
based on CAK75,
in which the mass outflow rate of the line-driven stellar wind 
is investigated in detail.
Such a discrepancy of the initial values leads to 
the difference of the wind structure.
In \citet{Risaliti10},
since the matter is overionized by X-ray irradiation
at the vicinity of the black hole,
the line force is made powerless and fails to launch the wind. 
\citet{Murray95} also reported that the failed wind appears 
in the vicinity of the black hole via X-ray irradiation.
In our work, 
the matter is not highly ionized via the high density
(see $\rho_0$ profiles in Figure \ref{init_cond}).
However, high density reduces the force multiplier
and prevents the launching of the wind 
as we have mentioned in section \S \ref{results}.
Also, the outer part of the disk contributes more
to the total mass loss rate ($\propto \rho_0 v_0 r^2$) 
than the inner disk in \citet{Risaliti10},
since $\rho_0$ and $v_0$ are constant in the radial direction.
In contrast, $\rho_0$ as well as $v_0$ increase with the 
decrease of $r$, meaning that the disk loses more mass
from the inner part of the launching region.

Here, we note that the wind structure calculated by the present 
non-hydrodynamic method is similar to that calculated 
by hydrodynamic simulations of PK04. 
The opening angle of the wind, $\sim 60^\circ$, for PK04 
approximately coincides with our result, $\sim 50^\circ$.
The column density of the wind in the moderate or 
low ionization state is around 
$10^{24-25} \rm{cm}^{-2}$ ($\theta\sim 70^\circ$) in PK04,
and $10^{24-25} \rm{cm}^{-2}$ ($\theta\sim 60^\circ$) 
and $\gtrsim10^{25} \rm{cm}^{-2}$ ($\theta\gtrsim 70^\circ$) 
in our model. 
The wind velocity is much larger in our work than in PK04.
The outward velocity is $100,000\,\rm{km\,s}^{-1}$ in our work,
but PK04, on the other hand, showed around $10,000\, \rm{km\,s}^{-1}$ 
(the peak value is $20,000\, \rm{km\,s}^{-1}$).
Such a discrepancy is though to be caused by that
the disk wind is launched 
from more inner region in our model than in PK04.
In PK04, as well as \citet{Risaliti10}, 
the density just above the disk surface 
is set to be constant in the radial direction.
Thus, the density at the wind base is
lower in PK04 than in our model
at the inner region.
The less dense matter is almost fully ionized 
and is not accelerated by line force.
The failed wind region in PK04 is larger 
than that in our model.
The wind launched from the inner region
is effectively accelerated by strong UV radiation 
from the inner part of disk,
leading to higher velocity of the wind. 
Here we comment that,
when the initial density is set to be the same as that in PK04,
we can approximately reproduce the results of PK04,
e.g., the wind velocity of $10,000\, \rm{km\,s}^{-1}$
and the failed wind due to overionization.

\citet{Feldmeier99} investigated the line driven wind in
cataclysmic variables.
Their streamlines are almost in pararell with each other.
Such a feature is similar with out resutls.
However, they reported that the opening angle of the wind
is $10^\circ-30^\circ$, which is much smaller than 
that of our model and PK04, $50^\circ-60^\circ$.
Although the streamlines bends above the disk in our model,
straight lines are assumed in \citet{Feldmeier99}.
Also, the dilution of the radiative flux by the wind
(self-shielding) is not taken into consideration 
in \citet{Feldmeier99}.

\subsection{Future work}






Although axisymmetric hydrodynamic simulations 
of the line-driven wind are attempted by 
PK00 and PK04,
three-dimensional hydrodynamic approaches 
would be important future work.
Recently, the time variation of the absorption features
has been reported by \citet{Misawa07}.
This implies that the wind structure changes with time
and/or non-axisymmetric pattern exists in the winds.
Such problems would be resolved by 
three-dimensional hydrodynamic simulations.
\citet{OP99}
have reported using one-dimensional calculations that
the line-driven wind exhibits violent density fluctuations
and many density peaks.
In addition,
\citet{Proga00} and PK04 have shown 
that high density blobs are sometimes generated in the wind.
However, their work is a two-dimensional study.
We will explore the three-dimensional simulations to investigate
the time evolution and non-axisymmetric structure of the winds.

In the case that the mass outflow rate is
comparable to the mass accretion rate of the disk,
we should solve the wind and disk, self-consistently.
The wind extract the mass, angular momentum, and the energy 
from the disk.
The disk structure, thus, changes via launching of the wind,
and the wind structure is subject to influence of the 
changing the disk structure.
As we have discussed above, 
the mass outflow rate is comparable to or larger than 
mass accretion rate for $\varepsilon=0.05-0.1$ in our model.
By X-ray observations of AGNs, \citet{Tombesi12} reported that 
the mass outflow rate via the ultra fast outflows
is comparable to the mass accretion rate onto the black hole.


In the present study,
we verify the line-driven wind model
based on the ionization parameter,
the outward velocity, and the column density.
However, we should calculate the emergent spectra
and directly compare these with the observations.
Such a study was recently attempted by 
\citet{Sim10}, in which 
they performed 
Monte Carlo radiative transfer simulations in the X-ray band.


\section{Conclusions}
We have studied the structure of the disk wind 
driven by the radiation force including the line force,
by calculating trajectories of the fluid elements which are launched 
from the surface of the geometrically thin and optically thick disk.
Here, the density and velocity at the wind base are set 
so as meet the condition of line-drive stellar wind (CAK75).
We have solved the equation of motion,
coupling with the mass conservation along the streamlines.
The radiation force, the ionization parameter, and the temperature
are calculated by taking into consideration
the extinction of the X-ray from the vicinity of the black hole
and the UV from the disk by the ejected matter.
We have compared the resulting wind structure with
the wind properties of X-ray observations of the BAL quasars
(the ionization parameter, the outward velocity, 
and the column density),
and estimated the probability (BAL probability),
with which the system is identified as a BAL quasar.



In the case that the Eddington ratio is $\varepsilon \sim 0.3-0.9$
and the black hole mass is $M_{\rm BH} \sim 10^{7-8.5}M_\odot$,
we found that the funnel-shaped disk wind 
with a opening angle of $\sim 50^{\circ}$ forms.
In this regime, since the wind shape is insensitive to the Eddington ratio 
and the black hole mass,
the BAL probability of several to $10$ percent does not change so much.
The observer's viewing angle is responsible for 
whether or not the quasars exhibit the broad absorption lines.
Our model is consistent with the phenomenological model 
of BAL quasars proposed by \citet{Elvis00}.

The wind is launched from the middle region of the disk, 
where the line force is strong enough to accelerate the matter.
At the inner region, since the larger initial density 
reduces the force multiplier (line force), 
the ejected matter immediately returns to the disk surface.
On the other hand,in the outer region, 
 since the UV radiation is attenuated via
the obscuration by the matter in the inner and middle regions,
the radiation force fail to launch the wind.
In the case of $M_{\rm BH} = 10^7M_{\odot}$ and $\epsilon = 0.3$,
the launching region is 
$30R_{\rm S}\lesssim r \lesssim 350R_{\rm S}$.

If the Eddington ratio is very small ($\varepsilon \lesssim 0.01$),
the disk wind does not appear,
since the UV intensity is too week for the matter 
to be accelerated by the line force.
Also, the disk wind is not launched 
when the black hole is too massive ($M_{\rm BH} \gtrsim 10^{9}M_\odot$).
This is because that 
the force multiplier at the wind base 
decreases with an increase of the black hole mass.
In addition, the disk temperature also decreases 
(UV emission region, $T_{\rm eff}>10^4$K, shrinks)
with an increase of the black hole mass.
Thus, for $\varepsilon \lesssim 0.01$ or 
for $M_{\rm BH} \gtrsim 10^{9}M_\odot$,
the BAL quasars are not observed
independently of the viewing angle.

\bigskip

We would like to thank Masahiro Morikawa and Yoshito Haba
for useful discussions.
This work is supported in part 
by the Ministry of Education, Culture, Sports, Science, and 
Technology (MEXT) Young Scientist (B) 20740115 (K.O.), 
21740150 (T. M.), 
and Scientific Research (C)
23540267 (K. W.), 
22540295 (H. S.).


\newpage


\newpage


\newpage
\end{document}